\documentclass[12pt]{article}
\begin{document}

\title{\bf  ${\bf SO(1,1)}$ dark energy model and the universe
transition
\thanks{ITP PHD Project
22B580; Liaoning Province Educational Committee Research Project;
National Nature Science Foundation Project of China.}}
\author{{\small Yi-Huan Wei$^{1,2}$, \quad Yu Tian$^{2}$}\\
{\small $^{1}$ Department of Physics, Bohai University, Jinzhou
121000, Liaoning, China}\\
{\small $^{2}$ Institute of Theoretical Physics, Chinese Academy
of Science, Beijing 100080, China}\\}
\maketitle
\begin{center}
{\bf Abstract}
\\ \vskip 0.5cm
\begin{minipage}{14cm}
\setlength{\baselineskip}{25pt plus1pt minus0pt} {\small ~~We
suggest the $SO(1,1)$ scalar field model of dark energy. In this
model, the Lagrangian may be decomposed as that of the real
quintessence model and the negative coupling energy term of $\Phi$
to $a$. The existence of the coupling term $L^c$ leads to a wider
range of $w_{\Phi}$ and overcomes the problem of negative kinetic
energy in the phantom universe model. We propose a power-law
expansion kinetics model of univese with time-dependent power,
which can describe the universe transition from ordinary
acceleration to super acceleration. We give also a simple
discussion of Big Rip singularity, and point out the possibility
that the universe driven by phantom avoid it.

PACS number(s)£º 98.80.Cq, 98.80.Hm
 }

 \vskip 0.5cm

\end{minipage}
\end{center}
\vspace{0.5cm}
\setlength{\baselineskip}{25pt plus1pt minus0pt}

\section {Introduction}
The observation of type Ia supernovae (SNe Ia) \cite{Riess} and
the cosmic microwave background power spectrum \cite {Ber} suggest
that our universe is undergoing accelerated expansion and
spatially flat, which have led to the consensus that in the
current universe consists of roughly 73\% dark energy and 27\%
matter. Probing the nature of the dark energy is one of the major
fundamental challenges in astronomy and physics today. In order to
explain the nature of dark energy, many models have been proposed,
such as, quintessence \cite{Quin}, k-essence \cite{K}, tachyon
\cite{T}, phantom \cite{Caldwell,Caldwel,P,Li}, Chaplygin gas
\cite{C}, spintessence \cite{Boyle},  etc.. So far, the
theoretical probe of dark energy focuses mainly on the evolution
of the dark energy density or the equation of state. The current
astronomical observations data can not determine completely the
nature of dark energy \cite{Mel,Alam}. The analysis based on the
SNe Ia data seems to favor the existence of a phantom energy in
the present universe \cite{Choudhury}. The phantom energy with
negative kinetic energy (KE) violates the null dominated energy
condition dynamically, the instability of the vacuum leads to the
vacuum decay \cite{Ca,Cl}. According to \cite{Ca}, the lifetime of
the phantom particle can exceeds cosmological time scale if the
momentum cutoff is smaller than $10^{-3}eV$. However, it is a pity
that there is no such energy scale in particle physics. Thus, the
problem of the instability of the vacuum is still an enigma, even
will be puzzle in a long time.

Another open question is in what domain the dark energy equation
of state ($w=\frac{p}{\rho}$) lies.  The SNIa data with the
constraints from WMAP observations rules out any rapid change in w
in recent epochs and are completely consistent with the
cosmological constant as the source of dark energy \cite{Jassal}.
However, combining the SNIa data with flat-universe constraints
including the cosmic microwave background and large-scale
structure, one can find $w =1.02^{+0.13}_{-0.19}$ (and $w < -0.76$
at the 95\% confidence level) for an assumed static equation of
state of dark energy \cite{Ries}. Clearly, it is very difficult to
obtain the severe constrain on $w$ from the current observations.
This is one of the main reasons why there are the various models
of dark energy. The models of dark energy may approximately be
classified into the following three possible categories: $w>-1$,
$w=1$ (cosmological constant) and $w<-1$ (phantom). Besides, one
can imagine that dark energy might have changed from the past
$w>-1$ to $w<-1$, since no result from the cosmological
observational data doesn't exclude this possible situation.
Providing that the above case, then such models that allow for an
arbitrariy $w$ will be needed.

Boyle, Caldwell and Kamionkowski proposed the spintessence model
for dark energy and dark matter \cite{Boyle}, which has a $U(1)$
symmetry and generalizes the quintessence model. From the
Lagrangian $L=\frac12(\dot{\phi}\dot{\phi}^*)-V(|\phi|)$ with
$\phi=\phi_1+i\phi_2$, $\phi^*=\phi_1-i\phi_2$ and
$|\phi|=\sqrt{\phi\phi^*}$, one can have the equivalent form with
a $SO(2)$ symmetry,
$L=\frac12({\dot{\phi}_1}^2+{\dot{\phi}_2}^2)-V(|\phi|)$. Compared
to the quintessence model with a real field, the spintessence
model make an important improvement due to the $U(1)$ symmetry. In
this paper, we propose the $SO(1,1)$ model of dark energy, in
which the Lagrangian density may be decomposed as
$L=L_\Phi^Q+L^c$, with $L_\Phi^Q$ that of quintessence and $L^c$
the coupling Lagrangian density of $\Phi$ to $a$. The model shows
some new features, it can't only describe the phantom universe but
also the universe transition from ordinary acceleration to super
acceleration phase, in principle. The paper is organized as
follows. In Sec. II, we show the $SO(1,1)$ model and discuss the
two special cases, one is the case that KE is much larger than the
absolute of coupling energy (CE), and another can describe the
variation of $w$ through $w=-1$. We suggest also the power-law
expansion scale factor with time-dependent power, which has an
advantage for illustrating the second case. In Sec. III, we
discuss simply the problem of Big Rip, and give a brief summary of
the $SO(1,1)$ model.

\section{$SO(2;\eta)$ model of dark energy}

Based the spintessence model \cite{Boyle} and the extended complex
model for dark energy \cite{Wei}, we propose the following dark
energy model, the Lagrangian density of which is given by
\begin{eqnarray}
L=\frac12({\dot{\phi}_1}^2-\eta{\dot{\phi}_2}^2)-V(\sqrt{\phi_1^2-\eta\phi_2^2}),
\label{eq1}
\end{eqnarray}
where $\phi_1$, $\phi_2$ are spatially homogeneous scalar fields,
$\eta$ is a real parameter and $V$ is the potential. The
Lagrangian (\ref{eq1}) possesses clearly certain symmetry. By
writing
\begin{eqnarray}
L=\frac12 \dot{\Psi}^{T_\eta}\dot{\Psi}-V(\sqrt{\Psi^{T_\eta}
\Psi}), \label{eq2}
\end{eqnarray}
\begin{eqnarray}
\Psi^{T_\eta}=\Psi^TM(\eta), \quad \Psi^T=(\phi_1,\phi_2)^T, \quad
M(\eta)=diag(1,\eta), \label{eq3}
\end{eqnarray}
where "T" denotes the transpose of matrix, then one can see that
under the transformation
\begin{eqnarray}
\Psi'=M\Psi, \quad M=\left(\begin{array}{cc} c(\alpha;\eta)&s(\alpha;\eta)\\
s(\alpha;\eta)&\eta c(\alpha;\eta) \end{array}\right), \quad
c^2(\alpha;\eta)-\eta s^2(\alpha;\eta)=1, \label{eq4}
\end{eqnarray}
where $\alpha$ is a real "angular" parameter,
$c(\alpha;\eta)=\sum_{n=0}^{\infty}\frac{\eta^n}{(2n)!}\alpha^{2n}$
and
$s(\alpha;\eta)=\sum_{n=0}^{\infty}\frac{\eta^n}{(2n+1)!}\alpha^{2n+1}$
\cite{WYH}, the Lagrangian (\ref{eq2}) holds invariant. We call
this symmetry the $SO(2;\eta)$ symmetry, which includes the
$SO(2)$ and $SO(1,1)$ symmetries.

Defining the new field variables $\Phi$ and $\theta$ by
$\Phi=\sqrt{\phi_1^2-\eta\phi_2^2}$ and
$\tan(\theta;\eta)=\frac{s(\theta;\eta)}{c(\theta;\eta)}=\frac{\phi_1}{\phi_2}$,
then from (\ref{eq1}) we have
\begin{eqnarray}
L_\Phi=\frac12({\dot{\Phi}}^2-\eta\Phi^2\dot{\theta}^2)-V(\Phi).
\label{eq5}
\end{eqnarray}
For $\eta=-1$, (\ref{eq5}) yields the Lagrangian density
\cite{Boyle,Gao} (see also Refs.\cite{Bilic}). For a spatially
flat, isotropic and homogeneous universe consisting of the
dust-like matter and the dark energy originating from the
$SO(2;\eta)$ scalar fields, we have the Friedman equations
\begin{eqnarray}
H^2=(\frac{\dot{a}}{a})^2=\frac{8\pi G}{3}(\rho_m+\rho_\Phi),
\label{eq6}
\end{eqnarray}
\begin{eqnarray}
\frac{\ddot{a}}{a}=-\frac{4\pi G}{3}(\rho_m+\rho_\Phi+3p_\Phi),
\label{eq7}
\end{eqnarray}
and the motion equations of fields
\begin{eqnarray}
\ddot{\Phi}+3H\dot{\Phi}+\eta\dot{\theta}^2\Phi+V'(\Phi)=0,
\label{eq8}
\end{eqnarray}
\begin{eqnarray}
\ddot{\theta}+(2\frac{\dot{\Phi}}{\Phi}+3H)\dot{\theta}=0,
\label{eq9}
\end{eqnarray}
with
\begin{eqnarray}
\rho_\Phi=\frac12(\dot{\Phi}^2-\eta\Phi^2\dot{\theta}^2)+V(\Phi),
\label{eq10}
\end{eqnarray}
\begin{eqnarray}
p_\Phi=\frac12(\dot{\Phi}^2-\eta\Phi^2\dot{\theta}^2)-V(\Phi),
\label{eq11}
\end{eqnarray}
where $H=\frac{\dot{a}}{a}$ is the Hubble parameter, a dot and a
prime denote derivatives with respect to $t$ and $\Phi$,
respectively. Equation (\ref{eq9}) is independent of the parameter
$\eta$, and the solution is
\begin{eqnarray}
\dot{\theta}=\frac{c}{a^3\Phi^2}, \label{eq12}
\end{eqnarray}
where $c$ is a constant.

Decomposing the Lagrangian density (\ref{eq5}) into $L_{\Phi}^Q$
and $L^c$
\begin{eqnarray}
L_\Phi=L_{\Phi}^Q+L^c, \quad
L_{\Phi}^Q=\frac12{\dot{\Phi}}^2-V(\Phi), \quad L^c=-\frac12 \eta
c^2a^{-6}\Phi^{-2}, \label{eq13}
\end{eqnarray}
then (\ref{eq11}) and (\ref{eq12}) may be reformulated to
\begin{eqnarray}
\rho_\Phi=\rho_\Phi^Q+\rho_\Phi^c, \quad
\rho_\Phi^Q=\frac12\dot{\Phi}^2+V(\Phi), \quad
\rho_\Phi^c=-\frac12\eta\Phi^2\dot{\theta}^2 \label{eq14}
\end{eqnarray}
\begin{eqnarray}
p_\Phi=p_\Phi^Q+p_\Phi^c, \quad
p_\Phi^Q=\frac12\dot{\Phi}^2-V(\Phi), \quad
p_\Phi^c=-\frac12\eta\Phi^2\dot{\theta}^2, \label{eq15}
\end{eqnarray}
where $\rho_\Phi^c$ and $p_\Phi^c$ are the contributions to the
total energy density and pressure from $L^c$. Defining
$w_c=\frac{p_\Phi^c}{\rho_\Phi^c}$, then there is $w_c=1$, which
is independent of $\eta$. In the $SO(1,1)$ model, i.e., the case
of $\eta=1$, the equation of state is given by
$w=\frac{\frac12\dot{\Phi}^2-V(\Phi)+\rho_\Phi^c}{\frac12\dot{\Phi}^2+V(\Phi)+\rho_\Phi^c}$,
which shows clearly some new features: the existence of the
negative CE, which leads to the wide range of $w$ and the possible
$w<-1$ even though the KE holds always nonnegative. Provided that
$KE\geq |CE|$, then $w_\Phi\geq -1$; if $KE< |CE|$, then $w_\Phi<
-1$; when $KE+CE$ changes from a positive to a negative value,
$w_\Phi$ changes from $>-1$ to $<-1$.Thus, the $SO(1,1)$ model of
dark energy may allow for an arbitrary value of $w$, in principle.

In the following, we will focus on the second and third cases.
First, let us discuss the phantom case, i.e., $w<-1$. Assuming
that the equations of state of matter and dark energy, $w_m$ and
$w_\Phi$, and the fractions of matter and dark energy, $\Omega_m$
and $\Omega_\Phi$ satisfying $\Omega_m+\Omega_\Phi\simeq1$, are
varying slowly, then from equations (\ref{eq6}) and (\ref{eq7})
one can obtain $a\simeq(\alpha+\beta
t)^{2/3(1+\Omega_mw_m+\Omega_\Phi w_\Phi)}$ with $\alpha$ and
$\beta$ two constants and $w_\Phi<-\frac{1+3\Omega_m
w_m}{3\Omega_\Phi}$. Defining $\Sigma=\Omega_mw_m+\Omega_\Phi
w_\Phi$ and letting $\alpha=-a_m\Sigma$ and
$\beta=a_m(1+\Sigma)/t_m$ with $t_m$ a constant time and
$a_m=a(t_m)$, then we have
\begin{eqnarray}
a=a_m[-\Sigma+(1+\Sigma)(\frac{t}{t_m})]^ {2/3(1+\Sigma)},
\label{eq16}
\end{eqnarray}
\begin{eqnarray}
H=\frac{\dot{a}}{a}=\frac{2}{3[-\Sigma t_m+(1+\Sigma)t]}, \quad
\dot{H}=\frac{-2(1+\Sigma)}{3[-\Sigma t_m+(1+\Sigma)t]^2}.
\label{eq17}
\end{eqnarray}
Considering the matter component as the pressureless fluid, i.e.,
$p_m=0$, then it evolves according to $\rho_m=\rho_{m0}(a_0/a)^3$,
where $a_0=a(t_0)$ and $\rho_{m0}=\rho_{m}(t=t_0)$ with $t_0$ the
age of the universe. In this case, noting that
$\frac{\ddot{a}}{a}=H^2+\dot{H}$, from equations (\ref{eq6}),
(\ref{eq7}), (\ref{eq12}) and (\ref{eq17}) we obtain
\begin{eqnarray}
\frac12\dot{\Phi}^2-\frac12c^2\Phi^{-2}a^{-6}=-\frac12\rho_m+\frac{1+\Sigma}{12\pi
G[-\Sigma t_m+(1+\Sigma)t]^2}, \label{eq18}
\end{eqnarray}
\begin{eqnarray}
V=-\frac12\rho_m+\frac{1-\Sigma}{12\pi G[-\Sigma
t_m+(1+\Sigma)t]^2}.
\label{eq19}
\end{eqnarray}

For equation (\ref{eq18}), let us consider such a special case
that $KE\ll |CE|$. Letting $M=\rho_{m}a^3$ and
$N=-\frac{1+\Sigma}{6\pi Gt_m^2}a_m^{6}$, then from equation
(\ref{eq18}) we obtain
\begin{eqnarray}
\Phi \simeq c[Ma^3+N(a/a_m)^{6-3(1+\Sigma)}]^{-\frac{1}{2}}.
\label{eq20}
\end{eqnarray}
For late time evolution, there are $\Omega_\Phi\rightarrow1$,
$\Sigma\sim w_\Phi$, and equation (\ref{eq20}) reduces to
\begin{eqnarray}
\Phi\simeq cN^{-\frac{1}{2}}(a/a_m)^{\frac{3}{2}\gamma_\Phi-3}.
\label{eq21}
\end{eqnarray}
Defining $\gamma_\Phi=1+w_\Phi$ and $\rho^c=-c^2\Phi^{-2}a^{-6}$,
then there is approximately
\begin{eqnarray}
\rho^c\simeq-\frac{2M_P^2\gamma_\Phi
}{3t_m^2}(\frac{a}{a_m})^{-3\gamma_\Phi} \label{eq22}
\label{eq22},
\end{eqnarray}
where $M_P=1/\sqrt{8\pi G}$ is the reduced Planck energy. Defining
$\rho_k=\frac{1}{2}\dot{\Phi}^2$, then from equation (\ref{eq21})
we obtain
\begin{eqnarray}
\rho_k=\frac{3c^2(\gamma_\Phi-2)^2}{8\gamma_\Phi M_P^2}a^{-6}
\label{eq33}.
\end{eqnarray}
From equations (\ref{eq22}) and (\ref{eq33}), one see that the
condition $\rho_k\ll|\rho^c|$ may be guaranteed if
$a\gg[\frac{3c(\gamma_\Phi-2)t_m}{4\gamma_\Phi
M_P^2}]^2a_m^{-\frac{3\gamma_\Phi}{6-3\gamma_\Phi}}$ is satisfied.

In what follows, we discuss the possible evolution of dark energy
from $w>-1$ to $w<-1$. In order to accomplish this purpose, we
assume the following scale factor
\begin{eqnarray}
a\sim t^n, \label{eq23}
\end{eqnarray}
with $n$ a time-dependent power. From (\ref{eq23}), one has the
Hubble parameter and its first derivative with respect to time
\begin{eqnarray}
H=\dot{n}\ln t+\frac{n}{t}, \quad \dot{H}=\ddot{n}\ln
t+2\frac{\dot{n}}{t}-\frac{n}{t^2}, \label{eq24}
\end{eqnarray}
where a dot denotes the derivative with respect to time. Assuming
that $n$ has the form
\begin{eqnarray}
n=n_0+bt, \label{eq25}
\end{eqnarray}
where $n_0$ and $b$ are two constants, then from equation
(\ref{eq24}) we obtain
\begin{eqnarray}
H=b(\ln{t}+1)+n_0t^{-1}, \label{eq26}
\end{eqnarray}
\begin{eqnarray}
\dot{H}=bt^{-1}-n_0t^{-2}. \label{eq27}
\end{eqnarray}

Equation (\ref{eq27}) implies a critical time
$t_c=\frac{n_0}{b}>0$ when the transition from ordinary
acceleration ($\dot{H}<0$) to super acceleration expansion phase
($\dot{H}>0$) occurs  (we call this transition the super expansion
transition, compared to the transition from decelerated to
accelerated expansion). Considering the matter component as the
pressureless fluid, then from equations (\ref{eq6}), (\ref{eq7}),
(\ref{eq10}), (\ref{eq11}), (\ref{eq26}) and (\ref{eq27}),  we
obtain
\begin{eqnarray}
\dot{\Phi}^2+2L_\Phi^c=-\rho_{m0}\frac{t_0^{3(n_0+bt_0)}}{t^{3(n_0+bt)}}+
\frac{2\rho_0(n_0-b t)}{3H_0^2t^2}, \quad
L_\Phi^c=-\frac{1}{2}c^2\Phi^{-2}t^{-6(n_0+bt)},
\label{eq28}
\end{eqnarray}
\begin{eqnarray}
2V=-\rho_{m0}\frac{t_0^{3(n_0+bt_0)}}{t^{3(n_0+bt)}}+
\frac{2\rho_0(b t-n_0)}{3H_0^2t^2}+\frac{2\rho_0[b
t(\ln{t}+1)+n_0]^2}{H_0^2t^2}, \label{eq29}
\end{eqnarray}
where $\rho_0$ and $H_0$ are the total energy density and Hubble
parameter of the current universe. For the case of both $b$ and
$n_0$ being positive, the term
$-\rho_{m0}\frac{t_0^{3(n_0+bt_0)}}{t^{3(n_0+bt)}}$ will fall
faster and faster than $\frac{2\rho_0(n_0-b t)}{3H_0^2t^2}$ as $t$
increases, thus for $t\gg 1$ equations (\ref{eq28}) and
(\ref{eq29}) will reduce to
\begin{eqnarray}
\dot{\Phi}^2-c^2\Phi^{-2}t^{-6(n_0+bt)}=\frac{2\rho_0(n_0-bt)}{3H_0^2t^2},
\label{eq30}
\end{eqnarray}
\begin{eqnarray}
2V=\frac{2\rho_0(b t-n_0)}{3H_0^2t^2}+\frac{2\rho_0[b
t(\ln{t}+1)+n_0]^2}{H_0^2t^2}. \label{eq31}
\end{eqnarray}
In an enough small neighborhood of $t_c$, there is $bt_c-n_0\simeq
0$ and thus equations (\ref{eq30}) and (\ref{eq31}) reduce further
to
\begin{eqnarray}
\Phi^2\simeq 2c\int t^{-3(n_0+bt)}dt, \quad V\simeq\frac{\rho_0[b
t(\ln{t}+1)+n_0]^2}{H_0^2t^2}. \label{eq32}
\end{eqnarray}
Assuming a small $b$, then $bt$ is a slowly changing function of
$t$ and we can approximately have
$\Phi^2\simeq\frac{2c}{1-3(n_0+bt)}t^{-3(n_0+bt)+1}$.

In the $SO(1,1)$ model, the kinetic part $\frac{1}{2}\dot{\Phi}^2$
is extended to $K_{eff}=\frac{1}{2}\dot{\Phi}^2+\rho_\Phi^c$. The
slow-rolling in quintessence and slow-climbing conditions in
phantom model are replaced by $|K_{eff}|\ll V$, here. In this
case, the evolution of dark energy (or phantom energy) will depend
mainly on the evolution of potential. Thus, the discussions on the
evolution properties of quintessence or phantom should be valid
for the current $SO(1,1)$ model.

\section{Discussions}
In the previous section we propose the $SO(1,1)$ model of dark
energy and have considered the two special cases. This section
will devoted to a simple discussion on Big Rip singularity and the
instability of the model and give a summary of the $SO(1,1)$
model.

For a phantom universe described by the power-law scale factor
$a=a_m[-w_{\Phi}+(1+w_{\Phi})(\frac{t}{t_m})]^{2/3(1+w_{\Phi})}$
with $w_{\Phi}<-1$ a constant, one can almost be sure to infer the
occurrence of Big Rip \cite{Caldwell,Caldwel}. However, provided
that $w_\Phi$ evolves according to $w_\Phi=-1+O(t^{-n})$ with
$n>1$, then the universe may avoid the Big Rip and will stay on
the approximate de Sitter phase forever. So, the phantom energy
doesn't always lead to a Big Rip of universe. From equations
(\ref{eq28}) and (\ref{eq29}) (or (\ref{eq30}) and (\ref{eq31})),
one can have
$w_\Phi=\frac{\dot{\Phi}^2+2L_\Phi^c-2V}{\dot{\Phi}^2+2L_\Phi^c+2V}\simeq
-1-\frac{2(bt^{-1}-n_0t^{-2})}{3[b(\ln{t}+1)+n_0t^{-1}]^2}\simeq
-1-\frac{2}{3t(\ln{t}+1)^2}+O(t^{-2}[\ln{t}]^{-1})$ for $t\gg
t_c$. This is an example that the universe is driven by phantom
but evades a Big Rip singularity.

In the $SO(1,1)$ model dark energy falls into the two ranges,
$w_\Phi>-1$ and $w_\Phi<-1$ corresponding to $KE+CE>0$ and
$KE+CE<0$, respectively. Clearly, in the latter case the model
violates the weak energy condition, $\rho_\Phi+p_\Phi<0$, and thus
in this case dark energy has an instable property, like the all
other phantom models. As has been seen in Sec. II, this
instability may lead to a Big Rip singularity, the approximate de
Sitter phase or the other cases, which depend on the behaviors of
dark energy.

From equations (\ref{eq13})-(\ref{eq15}), one can see the scalar
model for dark energy with $SO(1,1)$ symmetry contains the scalar
quintessence model, the appearance of the coupling Lagrangian
density $L^c$ is the result of the $SO(1,1)$ symmetry. In this
model, the KE term holds always nonnegative for either $w>-1$ or
$w<-1$. The negative CE term plays a fundamental role, and
decreases generically with the growth of $a$ and approaches zero
when $a\rightarrow \infty$. For example, it is proportional to
$a^{-6}$ for a constant $\Phi$. To sum up, our dark energy model
has the following features, the wide range of the equation of
state $w$, the nonnegative KE and the existence of the CE term or
the $SO(1,1)$ symmetry.

\end{document}